\documentclass[11pt]{iopart}

\usepackage{iopams}
\usepackage{color}
\usepackage{graphicx}
\usepackage{cases}
\usepackage{makeidx}
\makeindex
\usepackage {graphicx}
\def\>{\right\rangle}
\def\<{\left\langle}
\def\be{\begin{equation}}
\def\ee{\end{equation}}
\def\ba{\begin{array}{lll}}
\def\ea{\end{array}}

\def\beq{\begin{eqnarray}}
\def\eeq{\end{eqnarray}}
\begin{document}

\title{Spectral noise for edge states at filling factor $\nu=5/2$}
\author{\bf{M Carrega}$^{1}$, D Ferraro$^{2,3,4}$ A Braggio$^{3}$, N Magnoli$^{2,4}$ and M Sassetti$^{2,3}$}
\address{$^1$ NEST, Istituto Nanoscienze-CNR, and Scuola Normale Superiore, I-56126, Pisa, Italy}
\address{$^2$ Dipartimento di Fisica, Universit\`a di Genova, Via Dodecaneso 33, 16146 Genova, Italy}
\address{$^3$ SPIN-CNR, Via Dodecaneso 33, 16146 Genova, Italy}
\address{$^4$ INFN, Sezione di Genova, Via Dodecaneso 33, 16146 Genova, Italy}

\begin{abstract}
  We present a detailed analysis of finite frequency noise for the
  $\nu=5/2$ fractional quantum Hall state in a quantum point contact
  geometry.  The results are obtained within the Pfaffian and
  anti-Pfaffian models.  We show that the behaviour of the coloured noise allows unambigously to discriminate among tunneling
  excitations with different charges. Optimal values of the external bias are found in order to emphasize the visibility of the noise peak associated with the tunneling of a 2-agglomerate, namely an excitation with charge double of the fundamental one. These correspond to the regime in which the bias is larger than the neutral modes cut-off frequency. The dependence on the temperature is also investigated in order to discriminate between the considered models.
\end{abstract}

\section{Introduction}
One of the more challenging and intriguing examples of strongly
correlated electrons system is represented by the Fractional Quantum
Hall (FQH) fluid \cite{DasSarma97}. In recent years this peculiar
state of matter has been subject of many theoretical and experimental
studies, leading to the observation of a large variety of states at
different values of the filling factor $\nu$ \cite{Tsui99}.  Among
them, the $\nu = 5/2$ state \cite{Willett87} has recently attracted
increasing attention also in view of its promising application to the topologically protected quantum computation
\cite{Nayak08}.  Several models were proposed in order to describe
this peculiar state \cite{Boyarsky09} ranging from Abelian models,
already considered for other states \cite{Wen95, Halperin93} to more
exotic non-Abelian descriptions such as the Pfaffian model
\cite{Moore91, Fendley07} and its particle-hole conjugate, the
anti-Pfaffian one \cite{Lee07, Levin07,Bishara08}.  Despite different statistical predictions \cite{Stern08}, all these models
have the common prediction of a fundamental excitation with charge
$e/4$ ($e$ the electron charge), the single-quasiparticle (qp).  At the bulk level, signature of this
value have been recently found in \cite{Yacoby11}. For edge states
this result was confirmed in a Quantum Point Contact (QPC) geometry \cite{Chang03}
allowing tunnelling of excitations between the edge of the Hall bar.
Here, measurements of the differential backscattering conductance
\cite{Radu08} confirmed the presence of an $e/4$ tunnelling charge.
Other experiments focused on the zero frequency shot noise through
the QPC, as already used for different filling factors
\cite{DePicciotto97, Saminadayar97, Chung03}, also confirmed the
$e/4$ charge as the relevant tunneling excitation \cite{Dolev08}.

Recent  experiments carried out at extremely low temperatures \cite{Dolev10},
showed a richer phenomenology with an evolution of the tunnelling
charge from $e/4$ to $e/2$, lowering the temperature \cite{Carrega11}.
Similar behaviour has been found also for other composite edge states
\cite{Chung03, Bid09}, indicating the possible common trend of tunneling excitations  with 
charge greater than the fundamental one at low energies \cite{Ferraro08, Ferraro10b}.

Despite these preliminary indications, there is still an open debate about which charge is involved in the tunneling through the QPC.  The main purpose of this paper is to shed light on this issue and to unambingously determine the charge excitation who is participating to tunneling. In order to answer this question we analyze the symmetrized finite
frequency noise \cite{Rogovin74} as a property complementary and alternative
 to the current and the zero frequency noise.  We will focus on Pfaffian and anti-Pfaffian models.

From the experimental point of view, measurements of coloured noise
have been carried out for a QPC in a 2D electron gas without magnetic field
\cite{Zakka07}.  Nevertheless, great efforts are devoted to extend the
observations to more interesting cases including the FQH regime.
Here, much of the relevant physics is at frequencies around the Josephson frequency
$\omega_{0}=e^{*}V/\hbar$, ($e^{*}$ the charge of the elementary
excitation) that is in the range of GHz for external
bias $V$ in the range of $\mu$V.  These values,
although high, should be  experimentally observable. 

From a theoretical point of view the coulored noise was investigated for the Laughlin
sequence \cite{Laughlin83}, and predicted to show a resonance at the
Josephson frequency \cite{Chamon95, Chamon96, Dolcini05, Bena07}.  Also
the filling factor $\nu=5/2$ has been considered, but only for the
fundamental charge contribution in the Pfaffian model, in order to
extract signatures of the non-Abelian statistics of the excitations
\cite{Bena06}. 

In this paper we discriminate among the different tunneling charges
involved in the transport, being their contributions resolved at different
frequencies.  We demonstrate, for both models,  the necessity to consider voltages 
larger than the neutral modes cut-off frequency, in order to efficiently detect tunneling of the 2-agglomerate. The Pfaffian and anti-Pfaffian models 
are also compared regarding their different behaviour in the temperature scaling of the peaks. 

The paper is divided as follow. Sec. \ref{transport} introduces the
two models for filling factor $\nu=5/2$ and the tunneling Hamiltonian
for the QPC geometry.  The relations between finite frequency noise
and tunneling rates are discussed with the evaluation of the tunneling
scaling properties.  Sec. \ref{results} is devoted to the analysis
of the results for the frequency dependent noise as a function of
external bias and temperature, focusing on the possibility to detect
tunneling excitations with charge higher than the fundamental one. In Sec. \ref{conclusion} we sum up the main results of the paper.

\section{Transport properties}
\label{transport}

\subsection{Model}
We consider the effective field theoretical description of edge states
at filling factor $\nu=5/2$ for the Pfaffian \cite{Moore91,
  Fendley07}, and the disorder dominated anti-Pfaffian models
\cite{ Lee07, Levin07}.  The Lagrangian density
$\mathcal{L}=\mathcal{L}_{\mathrm{c}}+\mathcal{L}_{\mathrm{n}}$, 
with  ($\hbar=1$) 
\be \mathcal{L}_{\rm{c}}=-\frac{1}{2\pi}
\partial_{x}
\varphi_{\rm{c}}\left(\partial_{t}\varphi_{\rm{c}}+v_{\rm{c}}
  \partial_{x} \varphi_{\rm{c}}\right)
\label{lagrangian_charged}
\ee and 
\be \mathcal{L}_{\rm{n}}=-i\psi \left(\xi\partial_{t}
  \psi+v_{\rm{n}} \partial_{x}\psi\right)-\frac{\alpha}{4 \pi}
\partial_{x} \varphi_{\rm{n}} \left(\xi\partial_{t} \varphi_{\rm{n}}
  +v_{\rm{n}} \partial_{x}\varphi_{\mathrm{n}}\right)\,,
\label{lagrangian_neutral}
\ee describes the Hall fluid at filling factor $\nu=1/2$ with two
completely filled inert Landau levels that represent the \emph{vacuum} of
the theory. The charged bosonic field $\varphi_{\rm{c}}(x)$ is related
to the electron number density
$\rho(x)=\partial_{x}\varphi_{\mathrm{c}}(x)/2 \pi$, while
$\varphi_{\rm{n}}(x)$ is a bosonic neutral field, and $\psi(x)$ a
neutral Majorana fermion \cite{Wilczek09}.  The parameter $\xi = \pm
1$ denotes the direction of propagation of neutral modes with respect
to the charged one. The two models differ in the neutral sector
$\mathcal{L}_{\mathrm{n}}$ only, with $\alpha=0$ and $\xi=1$ for the Pfaffian model (P), $\alpha=1$ and $\xi=-1$ for the anti-Pfaffian (AP) one.
The propagation velocities of the charged and neutral modes are
indicated with $v_{\mathrm{c}}$ and $v_{\mathrm{n}}$ respectively.
Numerical calculations carried out for the Pfaffian case suggest that
$v_{\mathrm{c}}>v_{\mathrm{n}}$ \cite{Hu09}. In accordance with the results  \cite{Levin07} obtained
within the disorder dominated phase of the anti-Pfaffian model we assume an equal
velocity $v_{\mathrm{n}}$ both for the bosonic and fermionic part of
the neutral sector.  We also introduce the energy bandwidths
$\omega_{\mathrm{c/n}}=a^{-1}v_{\mathrm{c/n}}$, with $a$ a finite
length cut-off.  In all the paper we will assume $\omega_{\rm c}$ as the
largest energy.  The quantization of the bosonic fields is given by
the commutators \be \left[\varphi_{\rm{c/n}}(x),
  \varphi_{\rm{c/n}}(y)\right]=i\pi \nu_{\rm{c/n}} {\rm{sgn}}(x-y)\,,
\ee with $\nu_{\rm{c}}=1/2$ and $\nu_{\rm{n}}=\xi $, while the
Majorana fermion commutes with both.

Operators destroying an excitation along the edge are~\cite{Fendley07, Levin07} 
\begin{eqnarray}
\Psi_{\rm P}^{(\chi,m)}(x)&\propto& \chi(x)
e^{i\frac{m}{2}\varphi_{\mathrm{c}}(x)}\nonumber\\
\Psi_{\rm AP}^{(\chi, m,
  n)}(x)&\propto& \chi(x)
e^{i\left[\frac{m}{2}\varphi_{\mathrm{c}}(x)+\frac{n}{2}\varphi_{\mathrm{n}}(x)\right]}\,,
\label{Psi}
\end{eqnarray}
with $m,n \in \mathbb{Z}$ and where $\chi(x)$ belongs to the Ising conformal
field theory \cite{Ginsparg89}. The latter can be the identity operator $I$,
the Majorana fermion $\psi(x)$ or the spin operator $\sigma(x)$.  The
electric charge associated to the operators in (\ref{Psi}) is 
$(m/4)e$, and depends only on the charged mode. In the following,
we will call an $(m/4)e$-charged excitation as $m$-agglomerate
\cite{Carrega11, Ferraro08}.  The single-valuedness properties of 
$m$-agglomerates with respect to the electrons,
force $m$ and $n$ to be: even integers for $\chi=I$ or $\psi$, and odd
integers for $\chi=\sigma$ \cite{Bishara08}. Note that, in the Ising
sector, the spin operator $\sigma$, due to its non-trivial operator
product expansion $\sigma \times \sigma=I+\psi$ \cite{Ginsparg89},
leads to the non-Abelian statistics of the excitations \cite{Nayak08, Stern08}.  

\subsection{Tunneling through a QPC}
Tunnelling of a generic $m$-agglomerate between the two edges of the
Hall bar, through the QPC at $x=0$ is described by
$H^{(m)}_{T}=t_{m}\Psi^{(m)}_R(0){\Psi_L^{(m)}}^{\dagger}(0)+{\rm
  h.c.}$, where $\Psi^{(m)}_j$ is the annihilation operator of a
$m$-agglomerate (cf. Eq.(\ref{Psi})) on the right $j=R$ and left $j=L$
edges.  For simplicity, if not necessary, we will omit all other
indices present in (\ref{Psi}) treating the two models on the same footing. The tunnelling
amplitudes $t_{m}$ depend on factors like the geometry of the constriction, the intensity of the gate potential and the details of the edge
structure. Although in general these amplitudes can be energy dependent \cite{Chang03} for sake of clarity and to restrict the number of 
free parameters, in the following we will assume them as $m$-dependent constants.  Once a  
comparison with the experiments will become possible this assumption will be verified and possibly more complex choices will be considered. The total tunnelling Hamiltonian will consist of the sum over all possible excitations $H_T=\sum_{m}H_T^{(m)}$.

In this paper we will focus on the weak tunnelling regime. At lowest order in $H_T$ the
backscattering current of the $m$-agglomerate $\langle
I^{(m)}_{B}\rangle$ and the symmetrized noise $S_{B}^{(m)}(\omega)$
are directly written in terms of the tunneling rates \cite{Rogovin74}.
Here, we quote the final relations only, being these already
discussed in other works where full out of equilibrium approaches, such as the Keldysh Green's function method, were applied \cite{Rogovin74, Bena07, Bena06}. Using the detailed balance relation the current is \cite{Ferraro10a}
($k_{B}=1$) \be \langle I^{(m)}_{B}(\omega_0) \rangle=m e^{*}\left(1-
  e^{-m\omega_0/T}\right)\mathbf{\Gamma}^{(m)}( \omega_0)\,,
 \label{current_bal} \ee 
with the golden rule tunnelling rate at fixed voltage
\be
\mathbf{\Gamma}^{(m)}(E)=|t_{m}|^{2}
\int^{+\infty}_{-\infty} d\tau e^{im E\tau} G^{<}_{m,R}(0,-\tau) G^{>}_{m, L}(0,\tau)\,.
\label{tunn_rate}
\ee 
Here, $\omega_0 = e^* V$ ($e^{*}=e/4$) is the Josephson frequency associated to the single-qp with $V$ the bias. 
The correlators $G_{m,j}^{>}(0,t)=\langle \Psi^{(m)}_{j}(0,t)
\Psi^{(m) \dagger}_{j}(0,0)\rangle=(G^{<}_{m,j}(0,t))^*$ are the
two point Green's functions of the $m$-agglomerate operators on the edge $j=R,L$. It is worth to note that, at the lowest order in the 
tunnelling, the detailed balance relation is satisfied. Indeed, for weak tunneling the time between two tunneling events is much longer than any other 
time scale including the time that the collective excitations take to thermalize. In the Keldysh formalism one can easily show that the detailed balance 
condition derives directly, considering the contribution of the tunnelling at the lowest order.

Similarly, the noise spectral density 
\be
S_{B}^{(m)}(\omega)=\int_{-\infty}^{+\infty} d\tau \, e^{i \omega
  \tau}S_{B}^{(m)} (\tau)\,, 
  \ee 
  with  
  \be S_{B}^{(m)} (\tau ) =
\frac{1}{2} \left[ \langle I^{(m)}_B (\tau) I^{(m)}_B (0) \rangle +
  \langle I^{(m)}_B (0) I^{(m)}_B (\tau) \rangle - 2 \langle
  I^{(m)}_B(\tau)\rangle \langle I^{(m)}_B(0 ) \rangle \right]\,,
\label{symm_noise}
\ee
is related to the rate by extending what developed in Ref. \cite{Ferraro10a} to finite frequency. Here, for sake of simplicity, we focus on the backscattering 
current noise, whereas in experiments the total current noise is often
measured. At zero frequency, in the lowest order in the tunneling, the backscattering noise can be directly connected with the total current 
noise~\cite{Dolcini05, Ponomarenko99}. However, at 
finite frequency, one needs in general a four terminal setup and the backscattering noise is given by an appropriate 
measurement of the correlators of the different terminals. A complete and exhaustive study of these issues can be found in \cite{Bena07}. Our choice give us the possibility to keep our 
discussion as simpler as possible, directly in terms of a physically relevant quantity , the backscattering current. 
All the other possible correlators could be obtained  using the techniques developed in \cite{Bena07}. One then has 
\be
S_{B}^{(m)}(\omega)=\frac{(me^{*})^{2}}{2}\sum_{\varepsilon=\pm} \left[\mathbf{\Gamma}^{(m)}(\varepsilon\omega/m+ \omega_0 )+ \mathbf{\Gamma}^{(m)}(\varepsilon\omega/m-  \omega_0 )\right]\,,
\label{Noise_freq_rate}
\ee or equivalently in terms of the backscattering current \be
S_{B}^{(m)}(\omega)=\frac{1}{2}(me^{*})\sum_{\varepsilon=\pm}\coth{\left(\frac{\varepsilon\omega+
      m \omega_0 }{2 T}\right)} \langle I_B ^{(m)} (\varepsilon\omega/m+\omega_0 )\rangle\,.
\label{Noise_freq_curr}
\ee

This result is fully consistent with the non-equilibrium
fluctuation-dissipation theorem derived time ago for a single barrier tunnelling
Hamiltonian~\cite{Rogovin74, Dolcini05, Bena07}.  From
(\ref{Noise_freq_rate}) one can easily restore the well known result
for the zero frequency limit \cite{Ferraro10a, Martin04, Safi01}.

At the same perturbative order the total noise will be 
\be
S_{B}(\omega)=\sum_{m}S^{(m)}_{B}(\omega)\,.
\label{totalnoise}
\ee
being the contributions of the different $m$-agglomerate independent at the lowest order.  Note that the total  backscattering current  is also given by $\langle I_{B} \rangle=\sum_{m}\langle I^{(m)}_{B}\rangle$.

These simple relations (\ref{Noise_freq_curr}) and (\ref{totalnoise}) are due to the Poissonian statistics of the  tunnelling processes at lowest order in $|t_m|^2$ and to the independency of the sources of noise.

\subsection{Scaling behaviour}\label{scaling_sec}
Let us now focus on the evaluation of the rate (\ref{tunn_rate}),
starting with the zero temperature limit.  Here, the Green's functions
are \cite{Ginsparg89, Weiss99} 
\beq
\langle \chi(0,t) \chi(0,0) \rangle&=\frac{1}{(1+i \omega_{\mathrm{n}}t)^{\delta_{\chi}}},\qquad &\label{GFIsing}\\
\langle \varphi_{s}(0,t) \varphi_{s}(0,0) \rangle&=-|\nu_{s}| \ln{
  (1+i \omega_{s} t)} \qquad &s=\mathrm{c},
\mathrm{n}\label{GFbosonic} \eeq 
for the Ising sector and for the
charged and neutral bosonic fields respectively.  Due to the
independence of the Ising and bosonic sectors, the Green functions
factorize and the tunneling rate reads 
\be
\label{due_modi}
\mathbf{\Gamma}^{(m)} (E) = \frac{ |t_m|^2}{(2\pi a)^2} \int_{-\infty}^{+\infty}\!\!\!\! d t'  \frac{e^{ i m E t'} }{\left( 1 + i \omega_{\mathrm{c}} t'\right)^{ \frac{m^{2}}{4 }}}\frac{1}{\left( 1 + i \omega_{\mathrm{n}} t' \right)^{2 \delta_{\chi}+\alpha \frac{n^{2}}{2} }}\,,
\ee
with $\delta_{I}=0$, $\delta_{\psi}=1$ and $\delta_{\sigma}=1/8$ the conformal weights of the field in the Ising sector 
\cite{Nayak08,  Ginsparg89, Stern08}.  
The above integral can be performed leading to 
\beq
\mathbf{\Gamma}^{(m)}(E)&& =
\frac{ |t_m|^2}{2 \pi a^2} \left(\frac{m E}{\omega_{\mathrm{c}}}\right)^{\eta_m}\left( \frac{mE}{\omega_{\mathrm{n}}}\right)^{
\mu_{\alpha}} e^{-\frac{m E}{\omega_{\mathrm{c}}}} \frac{\left(m E\right)^{-1}}{\Gamma (\eta_m+\mu_{\alpha})} \nonumber \\
&& _1 F_1 \left[\mu_{\alpha}; \eta_m+\mu_{\alpha}; \left(\frac{m E}{\omega_{\mathrm{c}}} -\frac{m E}{\omega_{\mathrm{n}}}\right) \right]\Theta(m E)\,.
\label{risoluzione_due_modi}
\eeq Here,$\ _1F_1[a;b;z]$ is the Kummer confluent hypergeometric
function \cite{gradshteyn94}, $\eta_m=m^2/4$ the charge exponent and
$\mu_{\alpha}=2\delta_{\chi} +\alpha n^2/2$ the model dependent neutral
exponent ($\alpha=0$ for the Pfaffian model and $\alpha=1$ for the anti-Pfaffian one).

Note that for equal neutral and charge mode velocities, {\it{i.e.}} $\omega_{\mathrm{c}}=\omega_{\mathrm{n}}$, the above formula reduces to 
\beq
\label{limite_zeta_zero}
\mathbf{\Gamma}^{(m)}(E) =
\frac{|t_m|^2}{2 \pi a^2} \left(\frac{m E}{\omega_{\mathrm{c}}}\right)^{\eta_m +\mu_{\alpha}}\frac{(m E)^{-1}}{\Gamma (\eta_m +\mu_{\alpha})} e^{-\frac{mE}{\omega_{\mathrm{c}}}}\Theta(m E),
\eeq
recovering the standard result for a Luttinger liquid~\cite{Chamon96, Bena06, Braggio01, Cavaliere04, Cavaliere04b}.

The rate (\ref{risoluzione_due_modi}) shows different regimes
depending on the value of the energy with respect to the cut-off
frequency $\omega_{\rm n}$ of the neutral modes. Note that the charge cut-off $\omega_{\rm c}$ is assumed as the largest energy scale and it does not enter in the definition of these regimes. At low energies $E\ll\omega_{\rm n}$ the rate scales as \be
\mathbf{\Gamma}^{(m)}(E) \approx E^{4 \Delta_{\mathrm{P,AP}}^{(m)}-1} \ee
receiving contributions from both the charged and the neutral modes. In the
same limit, for frequencies close to the Josephson frequency $\omega\to
m\omega_0$, one has (cf. Eq.(\ref{Noise_freq_rate}))
\be
S_{B}^{(m)}(\omega\rightarrow m \omega_0)\approx (\omega- m \omega_0
)^{4 \Delta^{(m)}_{\mathrm{P,AP}}-1}\,.
\label{noiseJoseph}
\ee

Here, $\Delta_{\mathrm{P,AP}}^{(m)}$ are the $m$-agglomerate scaling
dimensions  of the operators
(\ref{Psi}), for the Pfaffian and anti-Pfaffian models respectively. They are deduced by the long-time behaviour at $T=0$ of the imaginary time
two-point Green's function $\langle T_{\tau}\Psi(\tau) \Psi^\dagger(0)\rangle
\propto |\tau|^{-2\Delta}$ \cite{Kane92} 
\be \Delta_{\rm
  P}^{(m)}=\frac{1}{2}\delta_{\chi}+\frac{1}{16} m^{2};\qquad
\Delta_{\rm AP}^{(m)}=\frac{1}{2}\delta_{\chi}+\frac{1}{16}
m^{2}+\frac{1}{8}n^{2}\,.
\label{Delta}
\ee
For the single-qp fundamental charge they are given by $m=1$, $n=\pm 1$ and $\chi=\sigma$
\be
\Delta_{\rm P}^{(1)}=\frac{1}{8};\qquad
\Delta_{\rm AP}^{(1)}=\frac{1}{4}\,.
\label{1_8}
\ee
For the next excitation, the $2$-agglomerate, the scaling are driven by the charged mode contribution only with 
$m=2$, $n=0$, $\chi=I$
\be \Delta_{\rm P}^{(2)}=\Delta_{\rm AP}^{(2)}=\frac{1}{4}.
\label{deltaagg}
\ee 
These behaviours indicate the single-qp particle as the most dominant excitation in the Pfaffian case, while  in the 
anti-Pfaffian case single-qp and 2-agglomerate have equal relevance with the same scaling dimensions~\cite{Levin07}.  
All other excitations, with higher charges, have higher scaling dimensions and can be safely neglected 
in the frequencies region $\omega\lesssim 2 e^* V$. 

Note that the peculiar values of the scaling (\ref{1_8}) and
(\ref{deltaagg}) implies a divergent power-law behaviour of the total
noise only for the Pfaffian case due to the single-qp contribution $S_{B}(\omega\rightarrow \omega_0)\approx (\omega-\omega_0)^{-1/2}$. In all
other cases this quantity appears flat due to the power-law cancellation in
(\ref{noiseJoseph}), therefore it will be more difficult to separate the
contribution of different excitations.  For example, at $\omega \rightarrow 2 \omega_{0}$ the 2-agglomerate can be detected only if $S_{B}^{(1)}(2\omega_0)<S_{B}^{(2)}(2\omega_0)$. As we will see, this condition can be
more easily achieved by considering energies (bias)
higher than the neutral mode cut-off, $\omega_0>\omega_{\mathrm{n}}$. Indeed,  in this case
the neutral modes are saturated modifying the scaling dimension (\ref{1_8}) of the
single-qp leading to a lower effective dimension \beq \tilde\Delta_{\rm
  P}^{(1)}=\tilde\Delta_{\rm AP}^{(1)}=\frac{1}{16}
\label{effscalingcomp}
\eeq
which will be active at the tail of the single-qp peak. On the other hands,
they leave unaffected the 2-agglomerate scaling as in
(\ref{deltaagg}). Then, the tail of the single-qp near $\omega=2\omega_0$ will drop faster with respect to the case when $\omega_0<\omega_ {\rm n}$
giving more visibility to the agglomerate itself. We will see in detail this behaviour in the next section.

At finite temperature the above behaviours will be smoothened with more remarkable changes near to the Josephson resonances $\omega=m\omega_0$ for $T>|\omega-m\omega_0|$.
To quantitatively determine the temperature influence on the  noise we consider  
the finite temperature  expressions for the Green's functions in
Eqs.(\ref{GFIsing}-\ref{GFbosonic}) both for the bosonic \cite{Ferraro08, Ferraro10a,
  Weiss99} and the Ising \cite{Fendley07} fields
  
\beq
\langle \chi(0,t) \chi(0,0) \rangle&=\left[\frac{|\Gamma\left(1+T/\omega_{\mathrm{n}}-i tT\right)|^{2}}{\Gamma^{2}\left(1+T/\omega_{\mathrm{n}}\right)(1+i \omega_{\mathrm{n}}t)}\right]^{\delta_{\chi}},\qquad &\label{GFIsingT}\\
\langle \varphi_{s}(0,t) \varphi_{s}(0,0) \rangle&=-|\nu_{s}| \ln{\left[\frac{\Gamma^{2}\left(1+T/\omega_{s}\right)
  (1+i \omega_{s} t)}{|\Gamma\left(1+T/\omega_{s}-i tT\right)|^{2}}\right]} \qquad &s=\mathrm{c},
\mathrm{n}\label{GFbosonicT} \eeq   

The tunneling rate (\ref{due_modi}) can be still explicitly evaluated
for temperatures lower than the bandwidths, namely
$T\ll\omega_{\mathrm{n}}, \omega_{\mathrm{c}}$, leading to 
\beq
\mathbf{\Gamma}^{(m)}(E) &=&\frac{|t_m|^2}{(2 \pi a)^2}\frac{ (2 \pi) ^{\eta_{m}+ \mu_{\alpha}}}{\omega_{\mathrm{c}}^{\eta_{m}} \omega_{\mathrm{n}}^{\mu_{\alpha}}}T^{\eta_{m}+\mu_{\alpha}-1}e^{\frac{mE}{2T}} \nonumber \\
 &\times& B\left(\eta_{m}+\mu_{\alpha}-i\frac{mE}{2 \pi T}; \eta_{m}+\mu_{\alpha}+i\frac{mE}{2 \pi T}\right)
\eeq
being $B(a; b)$ the Euler beta function \cite{gradshteyn94}. At higher temperatures  we evaluate the rate numerically.

Around the Josephson frequencies the noise in (\ref{Noise_freq_curr}) assume the simple form 
\be
S_{B}^{(m)}(\omega\rightarrow m \omega_0 )\approx TG_{B}^{(m)}(T)
\label{scaling_peaks}
\ee 
in terms of the $m$-agglomerate linear conductance 
\be
G_{B}^{(m)}(T)=(m e^{*})^{2} \frac{\mathbf{\Gamma}^{(m)}(0)}{T}.
\ee
The power-law behaviour for the height of the peaks as a function of the temperature is therefore given by $S_{B}^{(m)}(\omega\rightarrow m \omega_0 )\approx T^{\eta_{m}+\mu_{\alpha}-1}$ for $T\ll \omega_{\mathrm{n}}$ and  $S_{B}^{(m)}(\omega\rightarrow m \omega_0 )\approx T^{\eta_{m}-1}$ for $T\gg \omega_{\mathrm{n}}$.

\section{Results}
\label{results}
\begin{figure}
\centering
\includegraphics[width=0.60 \textwidth]{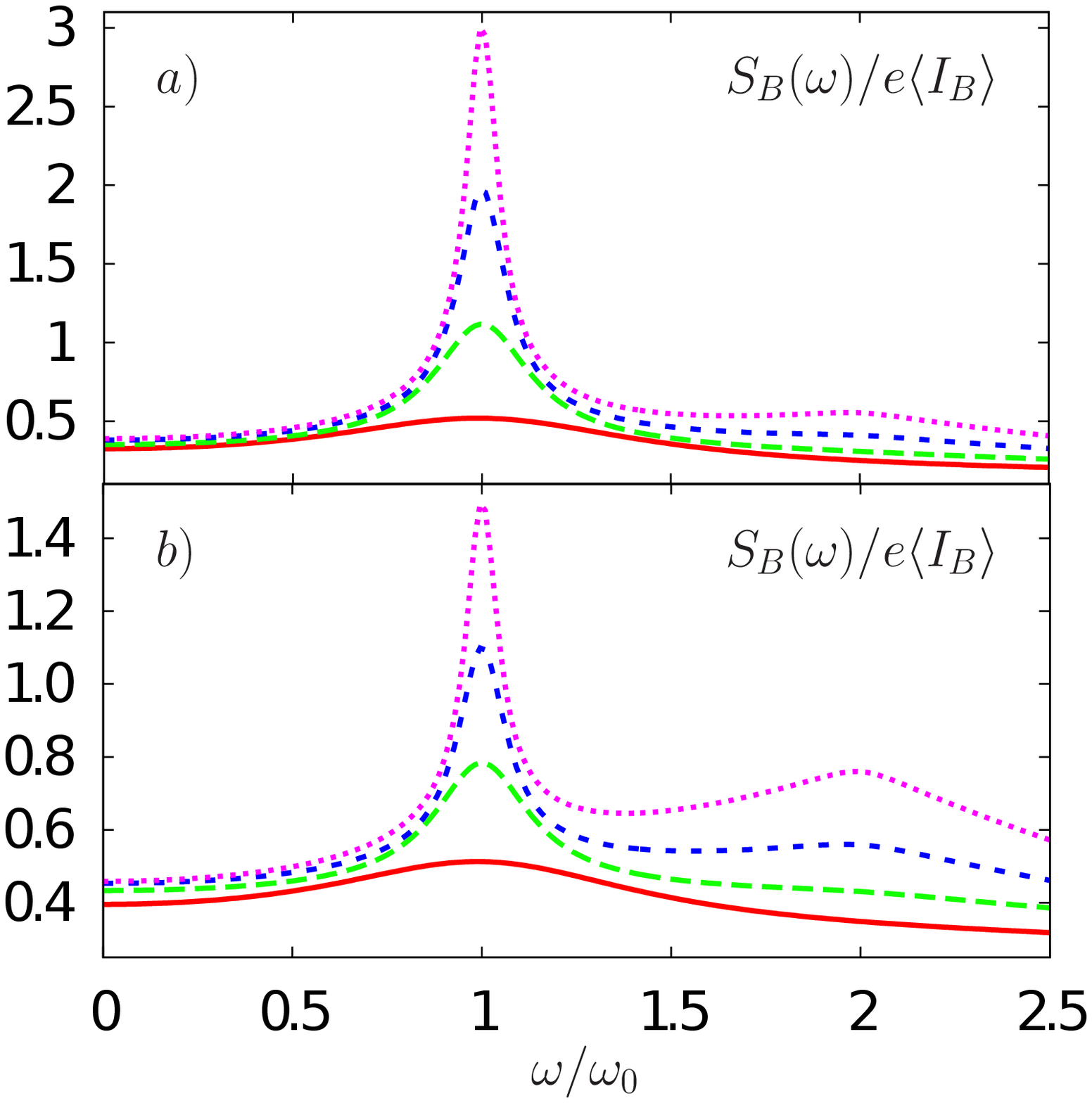}
\caption{Noise $S_{B}(\omega)/e \langle I_{B} \rangle$ as a function of frequency for the Pfaffian model with a) $\kappa=0.5$ and b) $\kappa=2$. The different values of the bias are:
  $\omega_0=30$ mK (red, solid), $\omega_0=100$ mK (green, long dash),
  $\omega_0=200$ mK (blue, short dash) and $\omega_0=300$ mK (magenta,
  dotted). Other parameters are: $\omega_{\mathrm{c}}=500$ mK,
  $\omega_{\mathrm{n}}=50$ mK, $T=10$ mK. Unit of frequency $\omega_{0}=eV/4$.}
 \label{Fig1}
\end{figure}

\begin{figure}
\centering
\includegraphics[width=0.6 \textwidth]{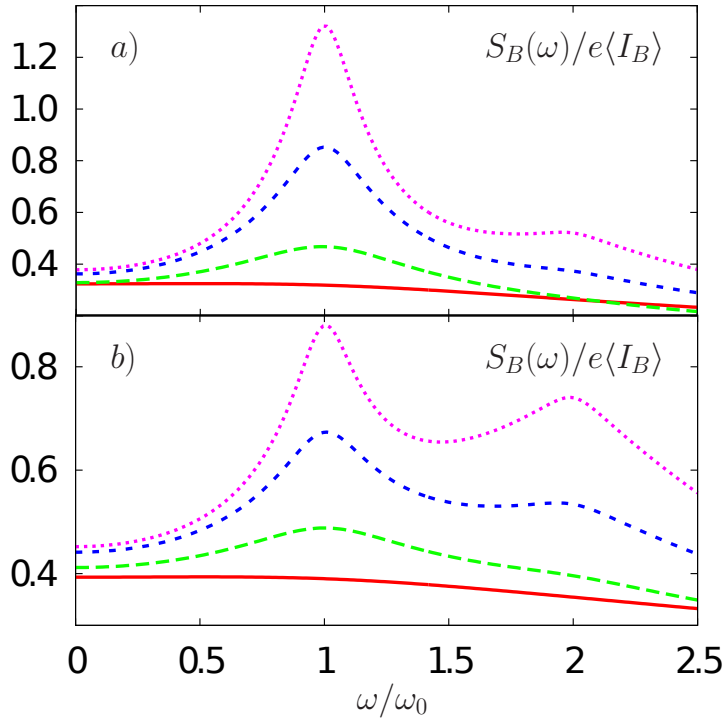}
\caption{Noise $S_{B}(\omega)/e \langle I_{B} \rangle$ as a function of frequency for the anti-Pfaffian model with a) $\kappa=0.5$ and b) $\kappa=2$. The different values of the
  bias are:  $\omega_0=30$ mK (red, solid), $\omega_0=100$ mK (green, long dash),
  $\omega_0=200$ mK (blue, short dash) and $\omega_0=300$ mK (magenta,
  dotted). Other parameters as in Fig.\ref{Fig1}. Unit of frequency $\omega_{0}=eV/4$.}
 \label{Fig2}
\end{figure}

In this Section we will discuss the behaviour of the ratio $S_{B}(\omega)/e
\langle I_{B} \rangle $ which, at zero frequency, corresponds to the
standard definition of the Fano factor  \cite{Ferraro08,
  Ferraro10a}.  We focus the analysis on a frequency regime up to the second Josephson
resonance $\omega=2\omega_0$, despite one could extend to higher
frequencies.  Therefore, we restrict the sum in the total noise
(\ref{totalnoise}) to $m=1,2$. For notational convenience we introduce
$\kappa=|t_{2}|^{2}/|t_{1}|^{2}$, which represents the relative
weight of the $2$-agglomerate amplitude in comparison to the single-qp
one. The charge mode bandwidth
is assumed of the same order of the activation gap for the Hall fluid
at $\nu = 5/2$ ($\omega_{\mathrm{c}} = 500$ mK) \cite{Xia04}.

Let us start by considering the noise as a function of frequency
varying the external bias $\omega_0$, for different ratios
$\kappa$. This will allow to find optimal values of $\omega_0$ necessary to 
increase the visibility of the 2-agglomerate. Notice that, for the higher values of the frequency $\omega$ plotted in the Figures 
one could expect also additional features related to the bulk quasiparticle dynamics, whose analysis is beyond the aim of this paper. The values of 
$\omega_0$ (i.e. the bias) must be carefully tuned considering that, the decreasing of this quantity leads to an increase of the relative importance of the thermal noise contribution. In our discussion we took the bias at quite high values to make the thermal effects on the finite frequency spectrum as smaller as possible. A
better compromise can be considered in the presence of a real experiments. Nevertheless, in such case, one need also to keep the frequency window of the measurement as smaller as possible and to consider energies smaller than the gap.

 Figures
\ref{Fig1}-\ref{Fig2} show this analysis for the two different models
(Fig. \ref{Fig1} for the Pfaffian mode and Fig.
\ref{Fig2} for the anti-Pfaffian one respectively).

All Figures show a general trend: the single-qp excitation peak at
$\omega=\omega_0$ is always well visible and much more pronounced in the Pfaffian model, while the 2-agglomerate peak  at
$\omega=2\omega_0$ increases notably its visibility in the regime
$\omega_0>\omega_{\mathrm{n}}$ namely at bias larger than the neutral cut-off
$\omega_{\mathrm{n}}$.  This result is in agreement with the discussion of the previous Section on the scaling
dimension of the excitations. Indeed, for both models the $2$-agglomerate noise contribution
is flat around  $\omega=2\omega_{0}$,  as can be verified from (\ref{noiseJoseph}) and from 
the scaling dimension $\Delta_{\rm P,\rm AP}^{(2)}=1/4$, and it decreases  
at higher frequencies as $(\omega-2 \omega_{0})/\omega_{\mathrm{c}}$, as a consequence of the exponential cut-off (cf. Eq. (\ref{risoluzione_due_modi})). It is then necessary to depress the tail of the single-qp contribution in order to emphasize its presence. This is indeed achieved for $\omega_0>\omega_{\rm n}$ (cf. Sec. \ref{scaling_sec}).

Another common feature is the behaviour as a function of the ratio $\kappa$ between the tunnelling amplitudes. By increasing  
$\kappa$ the height of the single-qp peak progressively decreases with a rising of the
2-agglomerate contribution in the high frequency region.  This is
connected to the reduction of the relative weight of the single-qp
noise with respect to the $2$-agglomerate. Indeed, for $\kappa=2$
(Figs.\ref{Fig1}b and \ref{Fig2}b) a much better resolved peak appears at
$\omega\approx 2 \omega_{0}$ whenever $\omega_0>\omega_{\rm n}$.

\begin{figure}
\centering
\includegraphics[width=0.60 \textwidth]{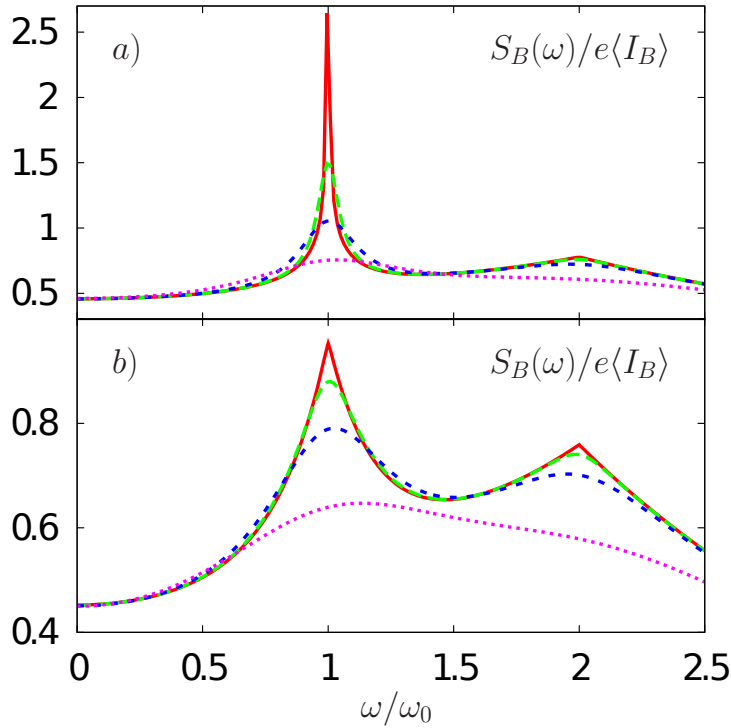}
\caption{Noise $S_{B}(\omega)/e \langle I_{B} \rangle$ as a function of frequency for the Pfaffian model a) and the anti-Pfaffian model b) at different temperatures: $T=1$ mK (red,
 solid), $T=10$ mK (green, long dashed), $T=30$ mK (blue,
  dashed) and $T=100$ mK (magenta, dotted). Other parameters are: $\kappa=2$, $\omega_{\mathrm{c}}=500$ mK, $\omega_{\mathrm{n}}=50$
  mK and $\omega_{0}=300$ mK. Unit of frequency $\omega_{0}=eV/4$.}
  \label{Fig3}
\end{figure}

\begin{figure}
\centering
\includegraphics[width=0.7 \textwidth]{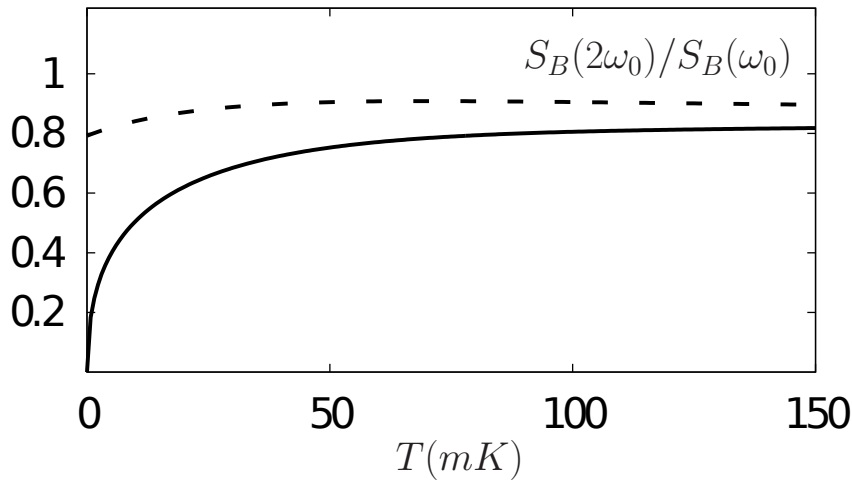}
\caption{Ratio $S_{B}(2\omega_0)/S_{B}(\omega_0)$ as a function of temperature for the Pfaffian model (black, full) and the anti-Pfaffian model (black, dashed). Other parameters are: $\kappa=2$, $\omega_{\mathrm{c}}=500$ mK, $\omega_{\mathrm{n}}=50$ mK and $\omega_{0}=300$ mK.}
\label{Fig4}
\end{figure}

Concerning the possible differences between the two models one can easily recognize that in the Pfaffian model the single-qp peak is more sharp and pronounced with respect to the one in the other.
The reason is again connected  to the scaling dimensions (cf. Eq.(\ref{1_8})) which are more relevant  in the first case. In order to better characterize this fact we consider the thermal evolution of the total finite frequency noise for both the models at $\kappa=2$. The results are shown in Fig. \ref{Fig3}. In the Pfaffian case (Fig. \ref{Fig3}a) the single-qp peak remains visible for a large range of temperatures despite the fact that the one of 2-agglomerate disappears quite soon.  For the other model (Fig. \ref{Fig3}b) the two peaks rounded in an analogous way by increasing the temperature. Note that for the higher considered temperature ($T=100$ mK) the described two-peak structure is completely washed out for both models. According to the power-law behaviours described in Sec. \ref{scaling_sec}, we expect for the ratio between the heights of the peaks $R=S_{B}(2\omega_0)/S_{B}(\omega_0)$ the power-law dependences on the temperature $R_{P}\approx T^{\frac{1}{2}}$ and $R_{AP}\approx constant$ for the Pfaffian and the anti-Pfaffian model respectively. These completely different trends clearly appear in Fig. \ref{Fig4}, where the ratios have been represented as a function of temperature and allow to better distinguish between the two models.

Before to conclude we would like to comment on the role of coupling with possible external environments.
As already discussed in several papers both for the Laughlin and Jain series and more generally for one dimensional Luttinger systems  \cite{Ferraro08, CastroNeto97, Sassetti94, Cuniberti97, Safi04, Rosenow02, Overbosch08, Yang03} these interactions cause a change in the scaling dimensions which could induce visible effects in transport properties. 

In particular, for the $\nu=5/2$ case the effects of the out of equilibrium $1/f$ noise determined by the coupling of the edge with  trapped charges in the substrate \cite{DallaTorre10} has been recently proposed as a source of interaction  for the charged and neutral  bosonic modes \cite{Carrega11}.

Despite the several theoretical predictions, no defined experimental results on the role of interaction were achieved until now,  especially for the $\nu=5/2$ case. This is the reason why we confined the analysis of the present work to the non-interacting case. However, the inclusion of interactions is quite straightforward according to the results described in the previous Section. 
The main effect is a dressing of the bosonic correlators in Eqs. (\ref{GFbosonic}) and (\ref{GFbosonicT}) 
$\langle \varphi_{s}(0,t) \varphi_{s}(0,0) \rangle_{\rm int}=g_{s}\langle \varphi_{s}(0,t) \varphi_{s}(0,0) \rangle$ ($s=\mathrm{c}, \mathrm{n}$)
with renormalization parameters $g_{\mathrm{c}}$  and $g_{\mathrm{n}}$ leading to the change in the scaling dimensions 

\be 
\Delta_{\rm  P}^{(m)}=\frac{1}{2}\delta_{\chi}+g_{\mathrm{c}}\frac{1}{16} m^{2};\qquad
\Delta_{\rm AP}^{(m)}=\frac{1}{2}\delta_{\chi}+g_{\mathrm{c}}\frac{1}{16}
m^{2}+g_{\mathrm{n}}\frac{1}{8}n^{2}\,.
\label{Deltarinorm}
\ee
For weak enough interactions the presented phenomenology is robust and we do not expect qualitative changes with respect to the above results. In the strong renormalized case two distinct regimes are possible. For $g_{\mathrm{c}}$ and $g_{\mathrm{n}}$ leading to scaling dimensions satisfying $1/4<\Delta_{\mathrm {P, AP}}^{(m)}<1/2$ the peaks at $\omega=m \omega_{0}$ turn into dips. Despite this change in the shape of the curve, it is still possible to resolve the contributions due to the different $m$-agglomerates. Conversely, when the interactions are so intense that $\Delta_{\mathrm {P, AP}}^{(m)}>1/2$ the total finite frequency noise increases monotonically with frequency and one loses track of the presence of the resonances. In this case it is more difficult to distinguish the presence of the various excitations. A detailed analysis of the renormalized case will be given elsewhere.

\section{Conclusion}
\label{conclusion}
In this paper we analyzed  the finite
frequency noise for  two credited theories proposed for the FQH
state at $\nu = 5/2$, the non-Abelian Pfaffian model and its particle-hole conjugate, the
anti-Pfaffian.  We considered the presence of the two most dominant excitations: the
single-qp, with charge $e/4$ and the $2$-agglomerate, with charge
$e/2$. The finite frequency noise has the unique possibility to resolve \emph{spectroscopically} the
contributions of the different excitations looking at different Josephson resonances. We showed that the peak associated to the $2$-agglomerate is more evident at Josephson frequency higher than the neutral mode cut-off, where the tail of the single-qp contribution decreases faster. We also considered the different evolutions of the height of the single-qp peak as a function of temperature, driven by the different scaling dimensions, as an important tool to discriminate between the models. 

\section*{Acknowledgements}
We thank T. Martin and F. Portier for useful discussions. We acknowledge the support of the EU-FP7 via ITN-2008-234970 NANOCTM.


\section*{References}
\begin{thebibliography}{10}
\bibitem{DasSarma97} Das Sarma S and Pinczuk A 1997 \emph{Perspective in quantum Hall effects: novel quantum liquid in low-dimensional semiconductor structures} (New York: Wiley)
\bibitem{Tsui99} Tsui D C 1999 Rev. Mod. Phys. \textbf{71} 891
\bibitem{Willett87} Willett R,  Eisenstein J P, Stormer H L, Tsui D C, Gossard A C and English J H 1987 Phys. Rev. Lett. \textbf{59} 1776
\bibitem{Nayak08} Nayak C, Simon S H, Stern A, Freedman M and  Das Sarma S 2008 Rev. Mod. Phys. \textbf{80} 1083
\bibitem{Boyarsky09} Boyarsky A, Cheianov V and Froehlich J 2009 Phys. Rev. B \textbf{80} 233302
\bibitem{Wen95} Wen X G 1995 Adv. Phys. \textbf{44} 405 
\bibitem{Halperin93} Halperin B I, Lee P A and Read N 1993 Phys. Rev. B \textbf{47} 7312
\bibitem{Moore91} Moore G and Read N 1991 Nucl. Phys. B \textbf{360} 362
\bibitem{Fendley07}  Fendley P, Fisher M P A and Nayak C 2007 Phys. Rev. B \textbf{75} 045317
\bibitem{Lee07} Lee S -S, Ryu S, Nayak C and Fisher M P A 2007 Phys. Rev. Lett. \textbf{99} 236807
\bibitem{Levin07} Levin M, Halperin B I and Rosenow B 2007 Phys. Rev. Lett. \textbf{99} 236806 
\bibitem{Bishara08} Bishara W,  Fiete G A and Nayak C 2008 Phys. Rev. B \textbf{77} 241306(R)
\bibitem{Stern08} Stern A 2008 Ann. Phys. \textbf{323} 204
\bibitem{Yacoby11} Venkatachalam V, Yacoby A, Pfeiffer L N and West K W 2011 Nature \textbf{469} 185
\bibitem{Chang03} Chang A M 2003 Rev. Mod. Phys. \textbf{75} 1449
\bibitem{Radu08} Radu I P, Miller J B, Marcus C M, Kastner M A, Pfeiffer L N and West K W 2008 Science \textbf{16} 899 
\bibitem{DePicciotto97} de Picciotto R, Reznikov M, Heiblum M, Umansky V, Bunin G and Mahalu D 1997 Nature \textbf{389} 162
\bibitem{Saminadayar97} Saminadayar L, Glattli D C, Jin Y and Etienne B 1997 Phys. Rev. Lett. \textbf{79} 2526 (1997)
\bibitem{Chung03} Chung Y C, Heiblum M and Umansky V 2003 Phys. Rev. Lett. \textbf{91} 216804
\bibitem{Dolev08} Dolev M, Heiblum M, Umansky V, Stern A and Mahalu D 2008  Nature \textbf{452} 829
\bibitem{Dolev10} Dolev M, Gross Y, Chung Y C, Heiblum M, Umansky V and Mahalu 2010 Phys. Rev. B. \textbf{81} 161303(R) 
\bibitem{Carrega11} Carrega M, Ferraro D, Braggio A, Magnoli N and Sassetti M 2011 Phys. Rev. Lett. \textbf{107} 146404
\bibitem{Bid09} Bid A, Ofek N, Heiblum M, Umansky V and Mahalu M 2009 Phys. Rev. Lett. \textbf{103} 236802 (2009)
\bibitem{Ferraro08} Ferraro D, Merlo M, Braggio A, Magnoli N and Sassetti M 2008 Phys. Rev. Lett. \textbf{101}, 166805
\bibitem{Ferraro10b} Ferraro D, Braggio A, Magnoli N and Sassetti M 2010 Phys. Rev. B \textbf{82} 085323
\bibitem{Rogovin74} Rogovin D and Scalapino D J 1974 Ann. Phys. \textbf{86} 1
\bibitem{Zakka07} Zakka-Bajjani E, Segala J, Portier F, Roche P and Glattli D C 2007 Phys. Rev. Lett. \textbf{99} 236803
\bibitem{Laughlin83} Laughlin R 1983 Phys. Rev. Lett. \textbf{50} 1395
\bibitem{Chamon95} Chamon C, Freed D E and Wen X G 1995 Phys. Rev. B \textbf{51} 2363
\bibitem{Chamon96} Chamon C, Freed D E and Wen X G 1996 Phys. Rev. B \textbf{53} 4033
\bibitem{Dolcini05} Dolcini F, Trauzettel B, Safi I and Grabert H 2005 Phys. Rev. B \textbf{71} 165309
\bibitem{Bena07} Bena C and Safi I 2007 Phys. Rev. B \textbf{76} 125317
\bibitem{Bena06} Bena C and Nayak C 2006 Phys. Rev. B \textbf{73} 155335
\bibitem{Wilczek09} Wilczek F 2009 Nature Physics \textbf{5} 614
\bibitem{Hu09} Hu Z -X, Rezayi E H, Wan X and Yang K 2009 Phys. Rev. B \textbf{80} 235330
\bibitem{Ginsparg89} Ginsparg P 1989 \emph{Applied Conformal Field Theory}, Les Houches Lectures, Ed. Brezin E and Zinn-Justin J Ed. (Amsterdam: North Holland)
\bibitem{Ferraro10a} Ferraro D, Braggio A, Magnoli N and Sassetti M 2010 New J. Phys. \textbf{12} 013012
\bibitem{Ponomarenko99} Ponomarenko V V and Nagaosa N 1999 Phys. Rev. B \textbf{60} 16865 
\bibitem{Martin04} Martin T 2005 \emph{Les Houches Session LXXXI}, ed. Bouchiat \emph{et al}, (Amsterdam: Elsevier)
\bibitem{Safi01} Safi I, Devillard P and Martin T 2001 Phys. Rev. Lett. \textbf{86} 4628
\bibitem{Weiss99} Weiss U 1999 \emph{Quantum dissipative system} (Singapore: World scientific)
\bibitem{gradshteyn94} Gradshteyn I S and Ryzhik I M 1994 \emph{Tables of integral, Series, and products} (London: Academic Press)
\bibitem{Braggio01} Braggio A, Sassetti M and Kramer B 2001 Phys. Rev. Lett. \textbf{87} 146802
\bibitem{Cavaliere04} Cavaliere F, Braggio A, Stockburger J T, Sassetti M and Kramer B 2004 Phys. Rev. B \textbf{70} 125323
\bibitem{Cavaliere04b} Cavaliere F, Braggio A, Stockburger J T, Sassetti M and Kramer B 2004 Phys. Rev. Lett. \textbf{93} 036803
\bibitem{Kane92} Kane C L and Fisher M P A 1992 Phys. Rev. Lett. \textbf{68} 1220
\bibitem{Xia04} Xia J S, Pan W, Vicente C L, Adams E D, Sullivan N S, Stormer H L, Tsui D C, Pfeiffer L N, Baldiwin K W and West K W 2004
  Phys. Rev. Lett. \textbf{93}, 176809
\bibitem{CastroNeto97} Castro Neto A H, Chamon C and Nayak C 1997 Phys. Rev. Lett. \textbf{79} 4629
\bibitem{Sassetti94} Sassetti M and Weiss U 1994 Europhys. Lett. \textbf{27} 311
\bibitem{Cuniberti97} Cuniberti G, Sassetti M and Kramer B 1997
J. Phys.: Condens. Matter \textbf{8} L21
\bibitem{Safi04} Safi I and Saleur H 2004 Phys. Rev. Lett. \textbf{93}, 126602 
\bibitem{Rosenow02} Rosenow B and Halperin B I 2002 Phys. Rev. Lett. \textbf{88} 096404
\bibitem{Overbosch08} Overbosch B J and Wen X G  arXiv:cond-math/0804.2087
\bibitem{Yang03} Yang K 2003 Phys. Rev. Lett. \textbf{91} 036802
\bibitem{DallaTorre10} Dalla Torre E G, Damler E, Giamarchi T and Altman E 2010 Nature Physics \textbf{6} 806 
\end{thebibliography}
\end{document}